\documentclass[twocolumn,english,aps,prb,twocolum,superscriptaddress,bibnotes,amsmath,amssymb,floatfix, dvipsnames]{revtex4-1}
\usepackage{float}
\usepackage[colorlinks=true,citecolor=blue,linkcolor=magenta]{hyperref}
\usepackage{multirow}
\usepackage[markup=nocolor, authormarkupposition=left]{changes} 
\usepackage{soul}
\usepackage[utf8]{inputenc}
\usepackage[english]{babel}
\usepackage{amsmath,amsfonts,amssymb}
\usepackage{mathrsfs}
\usepackage[T1]{fontenc}
\usepackage{url}
\usepackage{makecell}
\usepackage{amsmath}
\usepackage{amsfonts}
\usepackage{amssymb}
\usepackage{booktabs}
\usepackage{epstopdf}
\usepackage{graphicx}
\usepackage{float}
\usepackage{placeins}
\graphicspath{{./Figures/}}

\usepackage{changes}

\begin{document}

\title{Light coupling to photonic integrated circuits using optimized lensed fibers}

\author{Dengke Chen}
\affiliation{Shenzhen Institute for Quantum Science and Engineering, Southern University of Science and Technology, Shenzhen 518055, China}
\affiliation{International Quantum Academy, Shenzhen 518048, China}

\author{Zeying Zhong}
\affiliation{Shenzhen Institute for Quantum Science and Engineering, Southern University of Science and Technology, Shenzhen 518055, China}
\affiliation{International Quantum Academy, Shenzhen 518048, China}

\author{Sanli Huang}
\affiliation{International Quantum Academy, Shenzhen 518048, China}
\affiliation{Hefei National Laboratory, University of Science and Technology of China, Hefei 230088, China} 

\author{Jiahao Sun}
\affiliation{Shenzhen Institute for Quantum Science and Engineering, Southern University of Science and Technology, Shenzhen 518055, China}
\affiliation{International Quantum Academy, Shenzhen 518048, China}

\author{Sicheng Zeng}
\affiliation{Shenzhen Institute for Quantum Science and Engineering, Southern University of Science and Technology, Shenzhen 518055, China}
\affiliation{International Quantum Academy, Shenzhen 518048, China}

\author{Baoqi Shi}
\affiliation{International Quantum Academy, Shenzhen 518048, China}

\author{Yi-Han Luo}
\affiliation{International Quantum Academy, Shenzhen 518048, China}

\author{Junqiu Liu}
\email[]{liujq@iqasz.cn}
\affiliation{International Quantum Academy, Shenzhen 518048, China}
\affiliation{Hefei National Laboratory, University of Science and Technology of China, Hefei 230088, China}

\hyphenation{OVNAs}

\maketitle

\noindent\textbf{Efficient and reliable light coupling between optical fibers and photonic integrated circuits has arguably been the most essential issue in integrated photonics for optical interconnects, nonlinear signal conversion, neuromorphic computing,  and quantum information processing. 
A commonly used approach is to use inverse tapers interfacing with lensed fibers, particularly for waveguides of relatively low refractive index such as silicon nitride (Si$_3$N$_4$), silicon oxynitride, and lithium niobate. 
This approach simultaneously enables broad operation bandwidth, high coupling efficiency, and simplified fabrication. 
Although diverse taper designs have been invented and characterized so far, lensed fibers play equally important roles here, yet their optimization has long been under-explored. 
Here, we fill this gap and introduce a comprehensive co-optimization strategy that synergistically refines the geometries of the taper and the lensed fiber. 
By incorporating the genuine lensed fiber's shape into the simulation, we accurately capture its non-Gaussian emission profile, thereby nullifying the widely accepted approximation based on a paraxial Gaussian mode. 
We further characterize many lensed fibers and Si$_3$N$_4$ tapers of varying shapes using different fabrication processes. 
Our experimental and simulation results show remarkable agreement, both achieving maximum coupling efficiencies exceeding 80\% per facet.
Finally, we summarize the optimal choices of lensed fibers and Si$_3$N$_4$ tapers that can be directly deployed in modern CMOS foundries for scalable manufacturing of Si$_3$N$_4$ photonic integrated circuits.  
Our study not only contributes to light-coupling solutions but is also critical for photonic packaging and optoelectronic assemblies that are currently revolutionizing data centers and AI. }
 
\begin{figure*}[t]
\centering
\includegraphics{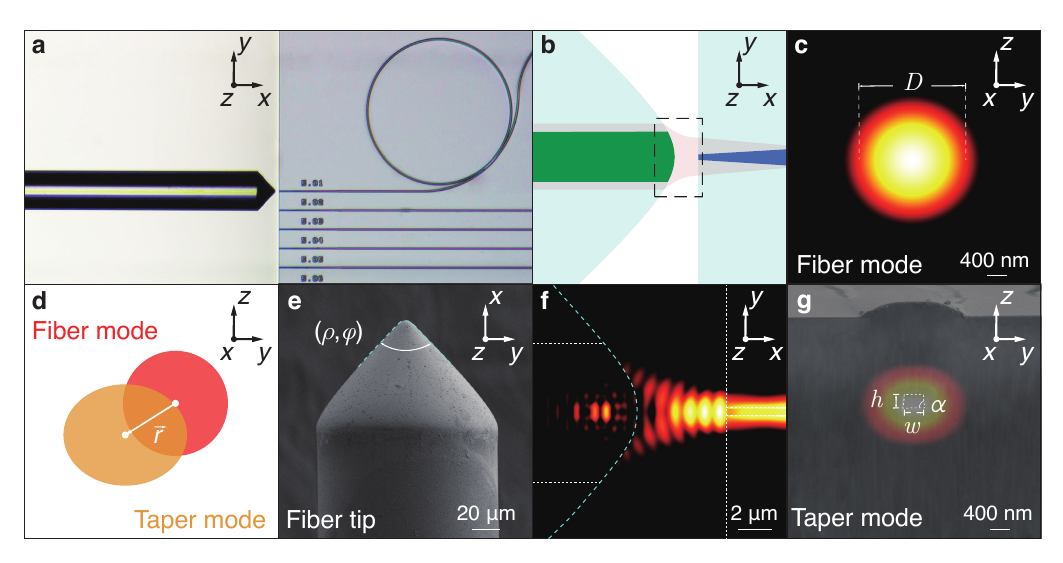}
\caption{
\textbf{Principle and schematic of light coupling from the lensed fiber to the Si$_3$N$_4$ taper}. 
\textbf{a}. 
Optical microscope image showing a lensed fiber edge-coupled to a Si$_3$N$_4$ chip, where a Si$_3$N$_4$ inverse taper is aligned to the fiber to receive light. 
\textbf{b}. 
Enlarged schematic of light coupling between the lensed fiber and the inverse taper. 
Doped fiber core, green. 
Si$_3$N$_4$ inverse taper, blue. 
Normal SiO$_2$ cladding, cyan.
Light field, transparent pink.
\textbf{c}. 
Simulated power distribution in logarithmic scale of the lensed fiber's focus mode. 
$D$, mode-field diameter.
\textbf{d}.
Principle of mode matching and overlap to calculate coupling efficiency $\eta_\text{ft}$. 
Accommodating one mode to the other and reducing misalignment vector $\vec{r}$ can improve $\eta_\text{ft}$. 
\textbf{e}.
SEM image of a lensed fiber's tip.
$\rho$, curvature radius.
$\phi$, conic angle.
\textbf{f}.
Simulated light propagation profile in logarithmic scale from the lensed fiber to the taper.
\textbf{g}.
SEM image of a subtractive taper's cross-section, overlapped with its TE eigenmode in logarithmic scale.
$w$, taper width.
$h$, taper height or thickness. 
$\alpha$, the taper's sidewall angle. 
}
\label{Fig:1}
\end{figure*}

\section{Introduction}
Photonic integrated circuits and waveguides utilize the unrivalled advantages of scalable CMOS manufacturing and have made significant progress over the last two decades.
They have enabled compact and cost-effective solutions in many applications, including telecommunication \cite{Agrell:16, Marin-Palomo:17}, datacenters \cite{Li:15, Raja:21}, neuromorphic computing\cite{Shen:17,Farmakidis:24}, LiDAR \cite{Poulton:17, Kim:21} and quantum information \cite{WangJW:20, Labonte:24,Larsen:25,Alexander:25}.
Currently,  integrated waveguides still often require fiber connection to other chips or devices such as isolators, amplifiers, and detectors.
The difference in cross-sectional and optical mode areas of fibers and waveguides is huge and can be up to thousands of times\cite{Son:18}.
Consequently, efficient and reliable light coupling between fibers and waveguides is pivotal and has long been the most essential topic in integrated photonics. 

In recent years, integrated photonics based on stoichiometric silicon nitride (Si$_3$N$_4$) has gained significant attention \cite{Moss:13, Xiang:22a}. 
This is due to that Si$_3$N$_4$ waveguides allow low optical loss, a wide transparency window from ultraviolet to mid-infrared, absence of two-photon absorption in the telecommunication band, moderate Kerr nonlinearity, and weak Raman and Brillouin gains \cite{Gyger:20}. 
Moreover, since Si$_3$N$_4$ has already been used as etch hardmasks, passivation layers, and stress layers in CMOS manufacturing of microelectronic circuits, its transformation into photonic circuits can naturally leverage the well-established scalable CMOS fabrication techniques \cite{Shekhar:24}.
Nevertheless, since Si$_3$N$_4$ is passive and amorphous, Si$_3$N$_4$ chips require a fiber connection to other active devices for light generation, modulation, and detection. 
Therefore, as mentioned earlier, achieving efficient and reliable fiber-chip coupling is equally crucial for Si$_3$N$_4$ waveguides.  
For example, the generation of microresonator-based Kerr frequency comb mandates sufficient on-chip power to trigger Kerr parametric processes \cite{Kippenberg:18,Pasquazi:18, Gaeta:19}.
In integrated neuromorphic networks, fiber-chip coupling loss limits the scalable depth of network layers and necessitates high laser power to maintain the computational signal-to-noise ratio, thereby significantly increasing energy consumption \cite{Yan:25,Farmakidis:24,Shen:17,Wangdl:24}.
For quantum information processing, fiber-chip coupling efficiency determines many metrics such as photon coincidence counts, the maximum squeezing level, and the fidelity of complex quantum states, ultimately influencing the scalability of fault-tolerant quantum systems\cite{Ding:25,Alexander:25,Larsen:25,Shen:25}.

Extensive effort has been made to optimize the light coupling from fibers to Si$_3$N$_4$ waveguides \cite{Marchetti:19}.
Although grating couplers are ubiquitously used in integrated photonics \cite{Taillaert:06}, they encounter severe challenges for Si$_3$N$_4$ waveguides with SiO$_2$ cladding, due to the low refractive index contrast\cite{Mak:18,Zhu:17} between the waveguide core ($n_\text{Si3N4}=1.99$) and the cladding ($n_\text{SiO2}\approx1.45$). 
Moreover, grating couplers suffer from narrow operation bandwidth, which impede their use in broadband nonlinear optical applications \cite{Masanovic:05,Laere:07, Roelkens:08}. 
An alternative to grating couplers is to use inverse tapers interfacing with lensed fibers, as shown in Fig. \ref{Fig:1}a. 
This approach can simultaneously achieve broadband operation, high efficiency, and simplified fabrication \cite{Almeida:03, Bakir:10, Liu:18}. 
In this configuration, the lensed fiber focuses the incident light to match the optical eigenmode of the waveguide taper, as shown in Fig. \ref{Fig:1}b, f.
The taper mode is weakly confined and thus features a large mode-field diameter (MFD).
When light propagates along the taper, the waveguide width slowly increases, adiabatically converging light to the tightly confined fundamental mode.
Together, these two components enable effective and reliable mode matching, as shown in Fig. \ref{Fig:1}d.
Various taper designs have been invented and characterized, including adiabatic parabolic tapers \cite{Almeida:03}, low-index converter encapsulation \cite{Shoji:02, McNab:03, Roelkens:05,Tsuchizawa:05,Bakir:10, Li:24}, and trident structures \cite{Hatori:14,He:20a, Liang:23,Gerguis:24}. 

In the meantime, despite the equally important roles of the lensed fibers, their optimization has long been under-explored. 
Conventional numerical simulations typically approximate the lensed fiber's focus mode as a paraxial Gaussian mode to avoid computational complexity. 
However, since the lensed fiber's tip has sub-wavelength structure, the actual mode profile can deviate from this approximation, manifesting a non-axial profile with a strong evanescent field\cite{Vaveliuk:07}. 
This discrepancy between the simulated and actual mode profiles compromises the prediction of coupling efficiency and hinders the development of reliable optimization strategies. 
Here, we fill this gap and introduce a comprehensive co-optimization strategy that synergistically refines the geometries of the taper and the lensed fiber. 
By incorporating the genuine lensed fiber's shape into the simulation, our approach reveals optimized designs with maximum coupling efficiencies exceeding 80\%, which agree well with experimental data. 

\begin{figure*}[t!]
\centering
\includegraphics{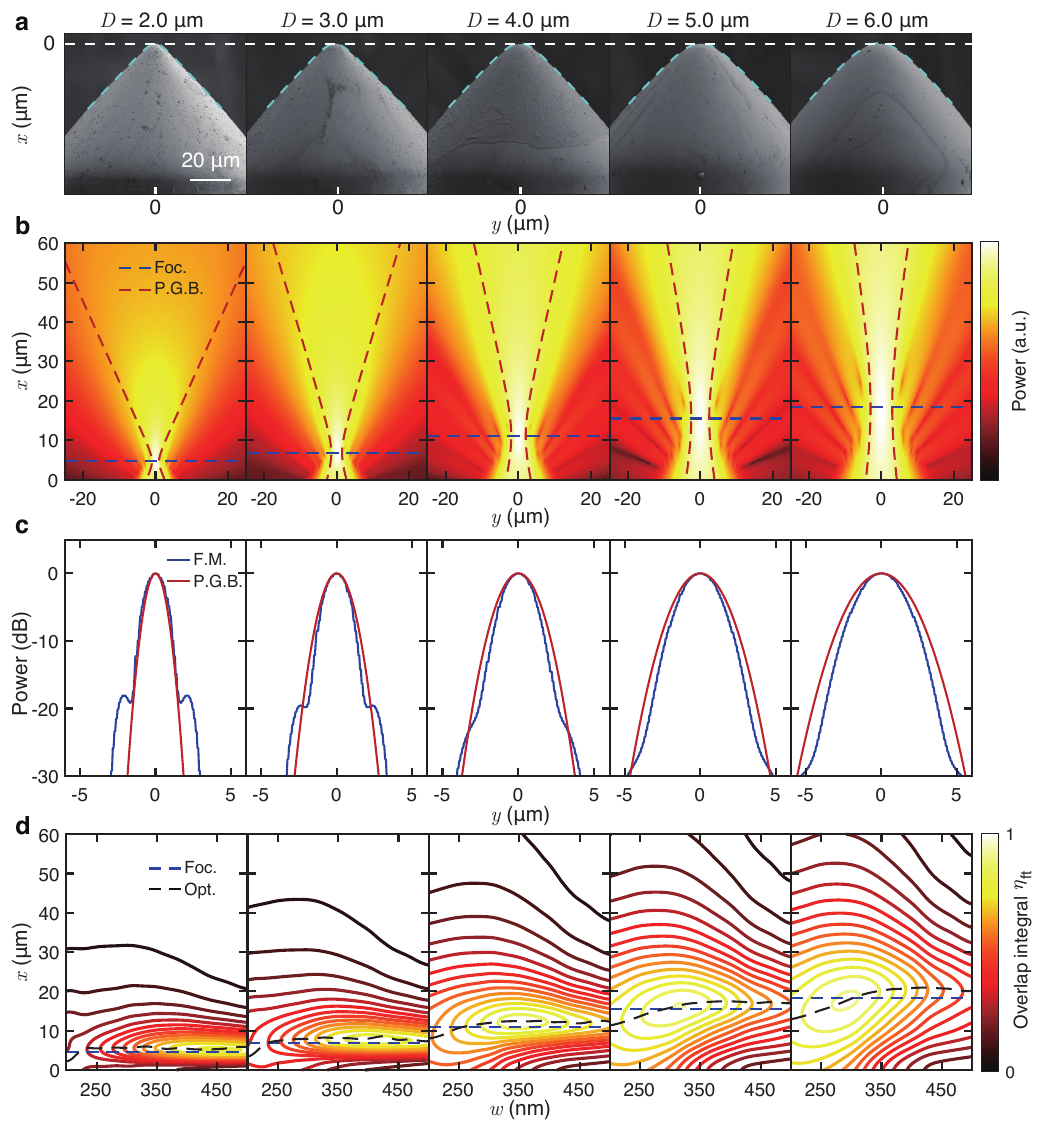}
\caption{
\textbf{Imaging, modeling and simulation of different lensed fibers. }
\textbf{a}. 
SEM images of lensed fibers' tips with $D=(2.0, 3.0, 4.0, 5.0, 6.0)$ $\mu$m. 
Cyan dashed curves outline the tips' hyperbolic shapes.
White dashed lines mark that the tips are placed at $x=0$. 
\textbf{b}.
Emission field in logarithmic scale of the lensed fibers.
Blue dashed lines mark the focal planes (Foc.) of the lensed fibers' emission modes. 
Red dashed curves mark the divergence of the paraxial Gaussian beam (P. G. B.). 
\textbf{c}.
Power distribution along the $y$ axis in the focal plane (blue dashed lines in \textbf{b}), for the lensed fiber’s focus mode (F. M.) and the paraxial Gaussian beam (P. G. B.) with different $D$ values. 
The comparison highlights discrepancy between the actual MFD extracted from the simulation and the labelled $D$. 
\textbf{d}.
Calculated $\eta_\text{ft}$ for lensed fibers of different $D$ values and tapers of varying $w$, over varying longitudinal misalignment along the $x$ axis. 
Blue dashed lines mark the focal planes (Foc.) of the lensed fibers' emission modes. 
Red dashed curves mark the optimal (Opt.) coupling position for maximum $\eta_\text{ft}$. 
As $D$ increases, $\eta_\text{ft}$ remains high over a broader range of $x$. 
As $D$ decreases, the red dashed lines approach the blue dashed lines. 
}
\label{Fig:2}
\end{figure*}


\section{Principle, modeling and numerical simulation}
The light coupling efficiency $\eta$ depends critically on the matching between the lensed fiber's focus mode and the inverse taper's mode.
As illustrated in Fig. \ref{Fig:1}d, higher similarity between their field profiles leads to greater coupling efficiency.
In addition, large mode profiles reduce the sensitivity to misalignment $\vec{r}$, thus enhancing spatial robustness. 
Therefore, key to efficient and reliable light coupling here is to engineer both modes' profiles via tailoring the geometries of the lensed fiber and the taper. 

The lensed fiber's focus mode is commonly characterized by its MFD, defined as the diameter $D$ at the $1/e^2$ power contour in the focal plane, as illustrated in Fig. \ref{Fig:1}c.
Under the \textit{assumption of a paraxial Gaussian beam}, $D$ can be determined from the measurement of far-field divergence angle $\theta$ as
\begin{equation}
D=\frac{4\lambda}{\pi \theta}
\label{DT}
\end{equation}
where $\lambda$ is the optical wavelength in air.  
Typically, vendors of commercial lensed fibers provide $D$ values for their products based on their measurements of $\theta$. 

Physically, $D$ is primarily governed by the hyperbolic curvature of the fiber tip, as shown in Fig. \ref{Fig:1}e. 
To accurately model lensed fibers, we perform scanning electron microscope (SEM) images of the tips to extract their profiles (see image processing techniques in Appendix A).
The extracted curve is fitted with a hyperbolic form
\begin{equation}
(\frac{x}{\rho}\tan^2{\frac{\phi}{2}}-1)^2-\frac{y^2}{\rho^2}\tan^2\frac{\phi}{2}=1,\quad x<0
\label{hyperbolic}
\end{equation}
where $\rho$ and $\phi$ are the curvature radius and the conic angle of the tip. 
In addition, $\phi$ is the angle between the asymptotes of Eq. \ref{hyperbolic}.

We purchase many commercial lensed fibers whose MFD values have been characterized by the vendor (Raysung Photonics Inc.), i.e. $D=(2.0, 3.0, 4.0, 5.0, 6.0)$ $\mu$m. 
Figure \ref{Fig:2}a shows the SEM images of these lensed fibers' tips, allowing the extraction of $(\rho,\phi)$ by fitting with Eq. \ref{hyperbolic}. 
In sum, the resulted $(D, \rho,\phi)$ values are (2.0 $\mu$m, 3.1 $\mu$m, 86$^\circ$), (3.0 $\mu$m, 4.4 $\mu$m, 88$^\circ$), (4.0 $\mu$m, 7.2 $\mu$m, 85$^\circ$), (5.0 $\mu$m, 10 $\mu$m, 84$^\circ$) and (6.0 $\mu$m, 13 $\mu$m, 80$^\circ$).

Next, we conduct a 3D finite-difference time-domain (FDTD) simulation to numerically analyze light propagation from a lensed fiber to an integrated Si$_3$N$_4$ waveguide at 1550 nm wavelength, as shown in Fig. \ref{Fig:1}b, f.  
We first model the lensed fiber's tip using the profile of Eq. \ref{hyperbolic} with fitted $(\rho,\phi)$ values. 
In the FDTD simulation, the tip is placed at $x=0$ and emits light into the free space $x>0$.
Figure \ref{Fig:2}b displays the simulated power distribution (normalized to its maximum) for the lensed fibers shown in Fig. \ref{Fig:2}a.
The blue dashed lines mark the focal planes of the lensed fibers obtained from the simulation, while the red dashed curves mark the $1/e^2$ power contour of a paraxial Gaussian beam (corresponding to Eq. \ref{DT}) with the same $D$. 
For each case, deviation is observed between the paraxial Gaussian beam profile and the actual profile of the lensed fiber's emission.
Figure \ref{Fig:2}c compares the power distribution along the $y$ axis in the focal plane (blue dashed lines in Fig. \ref{Fig:2}b), for the lensed fiber's focus mode (F. M., blue curves) and the paraxial Gaussian beam (P. G. B., red curves) with the given $D$. 
This comparison highlights the discrepancy between the actual MFD ($D_s$, extracted from the simulation) and the labelled MFD ($D$, provided by the vendors). 
Here, the values of $(D, D_s)$ are (2.0, 2.1), (3.0, 2.6), (4.0, 3.2), (5.0, 4.1), and (6.0, 5.0) $\mu$m, for the respective lensed fibers.
This observation nullifies the widely accepted assumption that a lensed fiber emits a Gaussian beam into free space \cite{Liu:18}. 
Consequently, an accurate FDTD simulation should account for the lensed fiber's actual shape. 

We then model the Si$_3$N$_4$ inverse taper \cite{Almeida:03, Liu:18}, which is placed at the chip facet to receive incident light from the lensed fiber and deliver the light to the multimode waveguide.  
In our experiment, the inverse tapers are fully embedded in SiO$_2$ cladding. 
The taper length is 300 $\mu$m, sufficiently long to adiabatically convert the light to the fundamental mode of the final multimode waveguide (of width $w>2$ $\mu$m). 
We fabricate these tapers using either the subtractive \cite{Gondarenko:09, Ji:17, Ye:23} or the additive processes \cite{Epping:15, Pfeiffer:16, Liu:21}. 
Details of our fabrication processes are presented in Appendices B and C. 
These two processes lead to two types of structures.
For \textit{subtractive} tapers, the waveguide thickness $h$ remains constant and is independent of the varying waveguide width $w$ (see illustration in Appendix B). 
Here, our fabricated subtractive tapers have fixed thickness ($h=320$ or $830$ nm) and sidewall angle ($\alpha=84^\circ$). 
Note that today Si$_3$N$_4$ waveguides of thickness around 300 and 800 nm are widely accessible in CMOS foundries as standard processes and services to the public.

In comparison, the \textit{additive} process results in tapers of varying $h$ and $w$, due to the aspect ratio-dependent etch effect (ARDE), as illustrated in Ref. \cite{Gottscho:92, Liu:18}.
Here, our fabricated additive Si$_3$N$_4$ waveguides have $h=710$ nm for $w>2000$ nm, as this thickness is ideal for creating near-zero anomalous group-velocity dispersion, critical for octave-spanning microcomb generation \cite{Pfeiffer:17, Anderson:22}.
We fabricate many additive tapers with different $w\in[170, 2800]$ nm, image them using SEM, and analyze these images. 
We empirically approximate the relation between ($h, \alpha, w$) as
\begin{equation}
\begin{cases}
h =\frac{k}{w}+b \\
\frac{h}{\tan\alpha} = k'w+b'
\end{cases}
\label{FIT}
\end{equation}
where $k=-1.473\times10^5$ nm$^2$, $b=770$ nm, $k'=9.8\times10^{-3}$, and $b'=-86$ nm are the fit parameters.
Details on the measurement and the fit are found in Appendix C. 

One step before running 3D FDTD simulation is to evaluate the overlap integral $\eta_\text{ft}$ as presented in Fig.~\ref{Fig:1}d. 
The integral $\eta_\text{ft}$ is defined as \cite{Solgaard:09}
\begin{equation}
\eta_\text{ft}(\vec{r})=\frac{\left|\int  \vec{E_\text{f}}(\phi,\rho,\vec{r}) \cdot \vec{E_\text{t}}^{*}(h,\alpha,w)\ \mathrm{d} A\right|^{2}}{\int\left|\vec{E_\text{f}}(\phi,\rho,\vec{r})\right|^{2} \mathrm{d} A \int\left|\vec{E_\text{t}}(h,\alpha,w)\right|^{2} \mathrm{d} A}
\label{OI}
\end{equation} 
where $\vec{E_\text{f}}$ and $\vec{E_\text{t}}$ denote the electric-field distribution of the lensed fiber's mode and the taper mode, 
$A$ represents the mode overlap area over the cross section in the $yz$ plane,  
and $\vec{r}$ specifies the misalignment between the two modes' centers. 
Thus, the calculation of $\eta_\text{ft}$ helps to evaluate the tolerance for mode misalignment $\vec{r}$. 

\begin{figure*}[t!]
\centering
\includegraphics{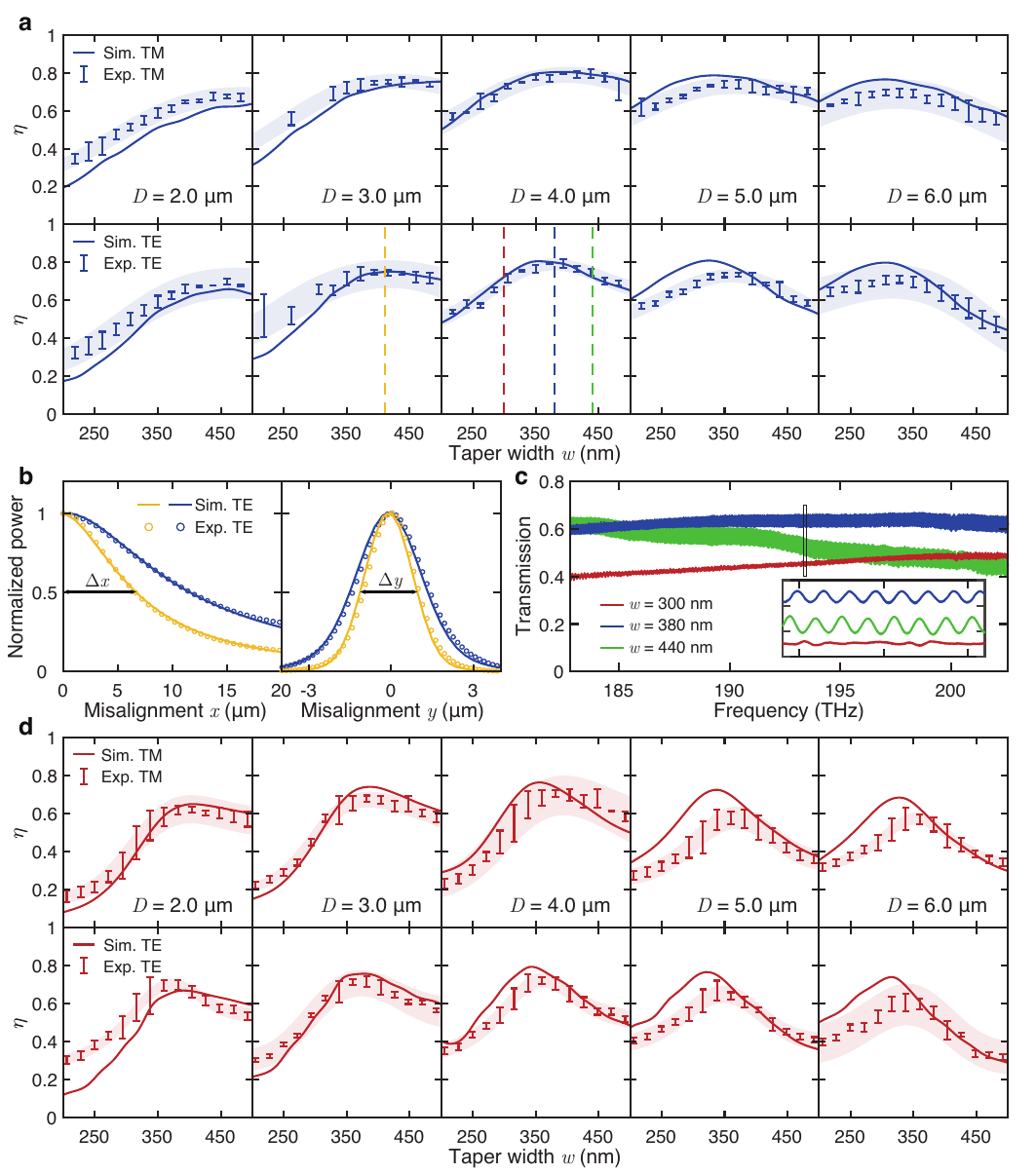}
\caption{
\textbf{Experimental characterization of light coupling efficiency $\eta$, misalignment tolerance, and transmission spectra}. 
\textbf{a, d}. 
Experimentally (Exp.) characterized $\eta$ for 320-nm-thick subtractive tapers (\textbf{a}) and 710-nm-thick additive tapers (\textbf{d}) with varying $D$ and $w$ under TE or TM polarization, in comparison with simulation (Sim.) results.
The error bars indicate twice the standard deviation for the three measurements of each taper with identical design parameters.
The vertical range of the shaded area indicates the 95\% confidence interval for $\eta$, as predicted by the Gaussian process regression model\cite{Wangj:23}.
\textbf{b}. 
Experimentally (Exp.) measured $\eta$ with 320-nm-thick subtractive tapers and TE polarization, for $D=3.0$ $\mu$m, $w=410$ nm (yellow data, also marked with yellow dashed line in \textbf{a}), and $D=4.0$ $\mu$m, $w=380$ nm (blue data, also marked with blue dashed line in \textbf{a}), in comparison with simulation (Sim.) results.
\textbf{c}.
Experimentally measured $\eta^2$ under TE polarization over the spectral range from 183 THz (1640 nm) to 203 THz (1480 nm), using $D=4.0$ $\mu$m and $w=300$, 380, and 440 nm (also marked with red, blue, and green dashed lines in \textbf{a}). 
Zoom-in shows the 15-GHz-FSR FP interference pattern. 
}
\label{Fig:3}
\end{figure*}

We investigate the dependence of $\eta_\text{ft}$ on the lensed fiber's $D$ and taper width $w$, under TE polarization. 
To simplify the calculation, in the simulation we align the two modes' centers in the $yz$ plane, i.e., $\vec{r}(x, y=0, z=0)$, which is experimentally easy. 
As an example, we consider subtractive tapers of $h=320$ nm.
Figure~\ref{Fig:2}d illustrates the calculated $\eta_\text{ft}$ for lensed fibers of $D=(2.0, 3.0, 4.0, 5.0, 6.0)$ $\mu$m and tapers of varying $w\in[200, 500]$ nm, with varying \textit{longitudinal misalignment} $\vec{r}(x, 0,0)$.
The red dashed curves mark the optimal coupling positions (for maximum $\eta_\text{ft}$), which generally lie near the lensed fibers' focal planes (blue dashed lines).
It is observed that as $D$ increases, $\eta_\text{ft}$ remains high over a broader range of $x$. 
Hence, lensed fibers with larger $D$ have a stronger tolerance to misalignment $\vec{r}$. 

Finally, we run 3D FDTD simulation and compute the overall coupling efficiency $\eta_{\text{sim}}$. 
In the simulation, for each lensed fiber of different $D$, we select the optimal $\vec{r}(x)$ based on the results in Fig.~\ref{Fig:2}d, and vary $w$.  
A typical simulated light propagation profile is shown in Fig.~\ref{Fig:1}f.
The standing wave patterns are due to light reflection in the boundaries. 
Details on the simulation are found in Appendix D. 
The simulated values $\eta_\text{sim}$ are compared with the experimentally measured data $\eta$ in Fig. \ref{Fig:3}, described in the next section. 

\vspace{0.3cm}
\section{Experimental results}
\begin{table*}[t!!]
\setlength{\tabcolsep}{6pt} 
\renewcommand{\arraystretch}{1.25} 
\centering
\belowrulesep=0pt
\aboverulesep=0pt
\begin{tabular*}{\textwidth}{@{\extracolsep{\fill}}c|c|ccc|ccc}
        \toprule
        \makecell{Process} & {($D$, $D_s$)} & \makecell{$w$\\(TE)} & \makecell{($\eta_{\text{ft}}$, $\eta_\text{max}$)\\ (TE)} &\makecell{($\Delta x$, $\Delta y$, $\Delta z$)\\ (TE)}& \makecell{$w$\\ (TM)} & \makecell{($\eta_{\text{ft}}$, $\eta_\text{max}$)\\ (TM)} &  \makecell{($\Delta x$, $\Delta y$, $\Delta z$)\\ (TM)} \\ 
        \midrule
        \multirow{5}*{\makecell{Subtractive \\320 nm thick}} 
        & (2.0, 2.1) $\mu$m &   470 nm & (70\%, 69\%) & (4.6, 1.6, 1.7) $\mu$m & 480 nm & (75\%, 68\%)  & (4.4, 1.6, 1.5) $\mu$m\\  
        & (3.0, 2.6) $\mu$m &   410 nm & (77\%, 74\%) & (7.2, 2.1, 2.0) $\mu$m & 430 nm & (85\% ,76\%)  & (6.8, 2.1, 2.0) $\mu$m\\ 
        & (4.0, 3.2) $\mu$m &   380 nm & (81\%, 80\%) & (9.6, 2.5, 2.6) $\mu$m & 400 nm & (88\%, 80\%)  & (8.9, 2.6, 2.6) $\mu$m\\ 
        & (5.0, 4.1) $\mu$m &   360 nm & (79\%, 73\%) & (9.4, 3.0, 2.8) $\mu$m & 370 nm & (84\%, 73\%)  & (9.3, 3.0, 2.7) $\mu$m\\ 
        & (6.0, 5.0) $\mu$m&   320 nm & (82\%, 70\%) & (14.8, 3.7, 3.2) $\mu$m & 330 nm & (84\%, 69\%)  & (13.5, 3.6, 3.2) $\mu$m\\ 
        \midrule
        \multirow{5}*{\makecell{Subtractive \\830 nm thick}} 
        & (2.0, 2.1) $\mu$m&   290 nm & (64\%, 65\%) & (3.8, 1.5, 1.5) $\mu$m & 220 nm & (70\%, 63\%)  & (3.9, 1.5, 1.4) $\mu$m\\ 
        & (3.0, 2.6) $\mu$m&   270 nm & (70\%, 69\%) & (6.7, 2.0, 1.9) $\mu$m & 180 nm & (78\%, 76\%)  & (6.9, 2.0, 1.9) $\mu$m\\ 
        & (4.0, 3.2) $\mu$m&   240 nm & (73\%, 68\%) & (9.7, 2.5, 2.5) $\mu$m & 160 nm* & (78\%, 80\%)  & (10.4, 2.6, 2.6) $\mu$m\\ 
        & (5.0, 4.1) $\mu$m&   230 nm & (66\%, 57\%) & (9.1, 3.0, 2.6) $\mu$m & 160 nm* & (66\%, 73\%)  & (9.3, 3.0, 2.6) $\mu$m\\ 
        & (6.0, 5.0) $\mu$m&   160 nm* & (82\%, 53\%) & (15.4, 4.3, 3.3) $\mu$m & 160 nm* & (58\%, 69\%)  & (14.5, 3.5, 3.0) $\mu$m\\ 
        \midrule
        \multirow{5}*{\makecell{Additive \\710 nm thick}} 
        & (2.0, 2.1) $\mu$m&   370 nm & (66\%, 70\%) & (5.0, 1.6, 1.5) $\mu$m & 410 nm & (73\%, 62\%)  & (4.1, 1.5, 1.4) $\mu$m\\ 
        & (3.0, 2.6) $\mu$m&   370 nm & (73\%, 71\%) & (7.1, 2.0, 1.9) $\mu$m & 400 nm & (76\%, 67\%)  & (6.6, 1.9, 1.8) $\mu$m\\ 
        & (4.0, 3.2) $\mu$m&   360 nm & (76\%, 72\%) & (9.6, 2.5, 2.4) $\mu$m & 400 nm & (69\%, 67\%)  & (9.0, 2.2, 2.3) $\mu$m\\ 
        & (5.0, 4.1) $\mu$m&   350 nm & (71\%, 65\%) & (8.5, 2.9, 2.5) $\mu$m & 370 nm & (68\%, 58\%)  & (8.3, 2.7, 2.4) $\mu$m\\ 
        & (6.0, 5.0) $\mu$m&   340 nm & (69\%, 60\%) & (14.9, 3.7, 3.0) $\mu$m & 350 nm & (70\%, 57\%)  & (13.5, 3.3, 2.8) $\mu$m\\      
        \hline
\end{tabular*} 
\caption{Summary of the optimal configurations for $\eta_\text{max}$.  
The asterisk $\ast$ marks the fact that 160 nm is currently the minimum dimension our photolithography can create for 830-nm-thick subtractive tapers. }
\label{Table}
\end{table*}

We experimentally characterize the coupling efficiency $\eta$ in the telecommunication band around 1550 nm wavelength, for both TE and TM polarizations.
The experimental setup is shown in Appendix E.  
The Si$_3$N$_4$ waveguides, each including two identical inverse tapers at the chip facets, are 5-mm-long.
Two lensed fibers of identical $D$ from the same vendor's batch are used to couple light into and out of a Si$_3$N$_4$ waveguide. 
Here we measure both the subtractive (320 and 830 nm thick) and the additive (710 nm thick) tapers, using five sets of lensed fibers of $D=(2.0, 3.0, 4.0, 5.0, 6.0)$ $\mu$m. 
Figure~\ref{Fig:3}a, d shows the measured $\eta$ with varying $w$, for 320-nm-thick subtractive tapers and 710-nm-thick additive tapers, while the data for 830-nm-thick subtractive tapers are shown in Appendix F and Figure~\ref{Fig:9}. 
The experimental data agree well with the FDTD simulation results, not only in trend but also in amplitude. 
This presents a significant improvement over the previous study \cite{Liu:18} that only shows agreement on the global trend but not the amplitude. 

For each case in Fig. \ref{Fig:3}a, d, an optimal $w$ is found for $\eta_\text{max}$ (i.e. the maximum $\eta$). 
We summarize the optimal sets of parameters in Table \ref{Table}, showing that
\begin{itemize}
\item $\eta_\text{max}$ initially increases and then decreases with increasing $D$.
\item Lensed fibers with a larger $D$ achieve $\eta_\text{max}$ with a smaller $w$.  
\item For a given $D$, $\eta_\text{max}$ can be different for the TE and TM polarizations due to the taper's anisotropy in cross-sectional geometry. 
\item Lensed fibers of $D=4$ $\mu$m seem to be the optimal choice, as the highest $\eta_\text{max}$ is achieved for both polarizations.
\end{itemize}

Next, we characterize the misalignment tolerance.  
We piezoelectrically displace the input lensed fiber along the $x, y, z$ axes around the optimal position.
The displacement is in 20 $\mu$m range with 125 nm per step.  
As examples, Figure \ref{Fig:3}b presents the measured $\eta$ with 320-nm-thick subtractive tapers and TE polarization, for two cases: 
$(D, w)$ = (3.0~$\mu$m, 410 nm) for the yellow data, and (4.0 $\mu$m, 380 nm) for the blue data.
Figure \ref{Fig:3}b shows the data with misalignment in the $x, y$ axes, and the data with misalignment in the $z$ axis are similar to those in the $y$ axis.
Moreover, numerical simulation based on the overlap integral Eq. \ref{OI} again agrees well with the experimental results.
Table \ref{Table} also summarizes the misalignment tolerance, where $(\Delta x, \Delta y, \Delta z)$ represents the maximum displacement for 3 dB reduction in $\eta$ from $\eta_\text{max}$ along the $x, y, z$ axes. 
It is found that $(\Delta x, \Delta y, \Delta z)$ mainly depends on $D$, which is consistent with the simulation result in Fig. \ref{Fig:2}d.

Finally, we evaluate the variation of $\eta$ over a wide spectral range from 183 THz (1640 nm) to 203 THz (1480 nm). 
Figure~\ref{Fig:3}c shows the experimentally measured $\eta^2$, to account for light into and out of the chip, passing through two lensed fibers and two tapers (see experimental setup in Appendix E). 
Here we use lensed fibers of $D=4.0$ $\mu$m coupled with 320-nm-thick subtractive tapers with $w=300$, 380 and 440 nm and TE polarization.
Light reflection on the chip's two facets causes a Fabry–Pérot (FP) interference pattern with approximately 15 GHz free spectral range (FSR), as shown in Fig.~\ref{Fig:3}c inset. 
The data show that a smaller $w$ results in a weaker FP interference. 
This is due to that smaller Si$_3$N$_4$ tapers have smaller effective refractive index $n_\text{eff}$, and the Fresnel reflection is $(n_\text{eff}-n_\text{air})^2/(n_\text{eff}+n_\text{air})^2$ with $n_\text{air}=1$ being the refractive index of air. 
For the optimal case with $w=380$ nm, $\eta$ is above 77\% with 3\%  variation over this spectral range. 

\vspace{0.3cm}
\section{Conclusion}
In conclusion, we have investigated--analytically, numerically, and experimentally--the light coupling between lensed fibers and Si$_3$N$_4$ inverse tapers of different parameters and configurations. 
Unlike extensive studies on the design and fabrication of inverse tapers, our study focuses more on the analysis of various lensed fibers, which is equally crucial but have long been overlooked. 
Specifically, our study involves accurate modeling of lensed fibers based on their high-resolution SEM images, comprehensive fine-mesh 3D FDTD simulations, and extensive experimental characterization of various fibers and tapers. 
We further compare our experimental data with simulation results, showing excellent agreement not only in trend but also in amplitude. 
Our study reveals a discrepancy in the actual emission mode of lensed fibers with the commonly accepted paraxial Gaussian beam. 
Moreover, we summarize the optimal choices of lensed fibers and Si$_3$N$_4$ tapers that can be directly deployed in modern CMOS foundries for scalable manufacturing of Si$_3$N$_4$ photonic integrated circuits. 
Our study not only contributes to light-coupling solutions in optical interconnects, nonlinear signal conversion, neuromorphic computing, and quantum information processing, but is also critical for photonic packaging and optoelectronic assemblies, which are currently revolutionizing datacenters and AI. 

\medskip
\begin{footnotesize}

\noindent \textbf{Acknowledgments}: 
We thank Zhenyuan Shang and Chen Shen for helping fabrication, Yue Hu for helping chip characterization, and Shuyi Li for fruitful discussion.
We acknowledge support from the National Natural Science Foundation of China (Grant No.12404417, 12261131503), 
Innovation Program for Quantum Science and Technology (2023ZD0301500), 
National Key R\&D Program of China (Grant No. 2024YFA1409300),
Shenzhen-Hong Kong Cooperation Zone for Technology and Innovation (HZQB-KCZYB2020050), 
Shenzhen Science and Technology Program (Grant No. RCJC20231211090042078), 
and Guangdong-Hong Kong Technology Cooperation Funding Scheme (Grant No. 2024A0505040008). 

\noindent \textbf{Author contributions}: 
D. C. performed the experiment and simulation, with the assistance and advices from S. Z., B. S. and Y.-H. L..
Z. Z., S. H. and J. S. fabricated and image the silicon nitride chips. 
J. L. supervised the project. 
D. C. and J. L. wrote the manuscript, with input from others.

\noindent \textbf{Conflict of interest}: 
J. L. is a founder of Qaleido Photonics, a start-up that is developing heterogeneous silicon nitride integrated photonics technologies. 
Others declare no conflicts of interest.

\noindent \textbf{Data Availability Statement}: 
The code and data used to produce the plots within this work will be released on the repository \texttt{Zenodo} upon publication of this preprint.

\end{footnotesize}

\vspace{0.3cm}
\section*{Appendix A: SEM image processing}
\begin{figure}[h!]
\centering
\includegraphics[width=0.5\textwidth]{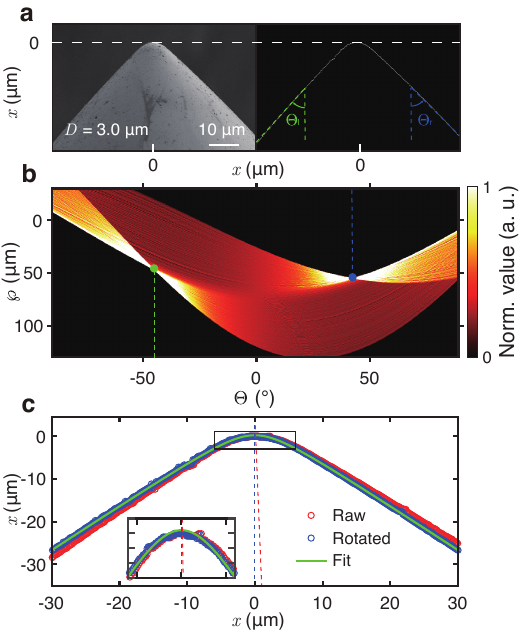}
\caption{
\textbf{SEM image processing.} 
\textbf{a.}
Edge extraction of the lensed fiber's tip.
Left panel, gray-scale SEM image.
Right panel, the corresponding edge profile obtained through sequential image processing, including $k$-means clustering, edge detection, and anomaly removal algorithms. 
The angles $\Theta_l$ and $\Theta_r$ define the fiber tip's geometry, representing the angles between its asymptotic lines and the $x$ axis. 
\textbf{b.}
Line detection via Hough transform. 
This panel visualizes the transformation of the edge image (from \textbf{a}, right) from the image space to its corresponding Hough parameter space.
The parameters $\wp$ (distance) and $\Theta$ (angle) characterize potential lines, with distinct peaks in this space indicating the most prominent linear segments.
\textbf{c.}
Geometric alignment and hyperbolic fit.
The solid green curve is the best-fit hyperbolic function applied to the raw edge coordinates after rotational correction.
The dashed red and blue lines represent the symmetry axes of the contour curves (raw and rotated data).
}
\label{Fig:4}
\end{figure}

\begin{figure*}[t!]
\centering
\includegraphics{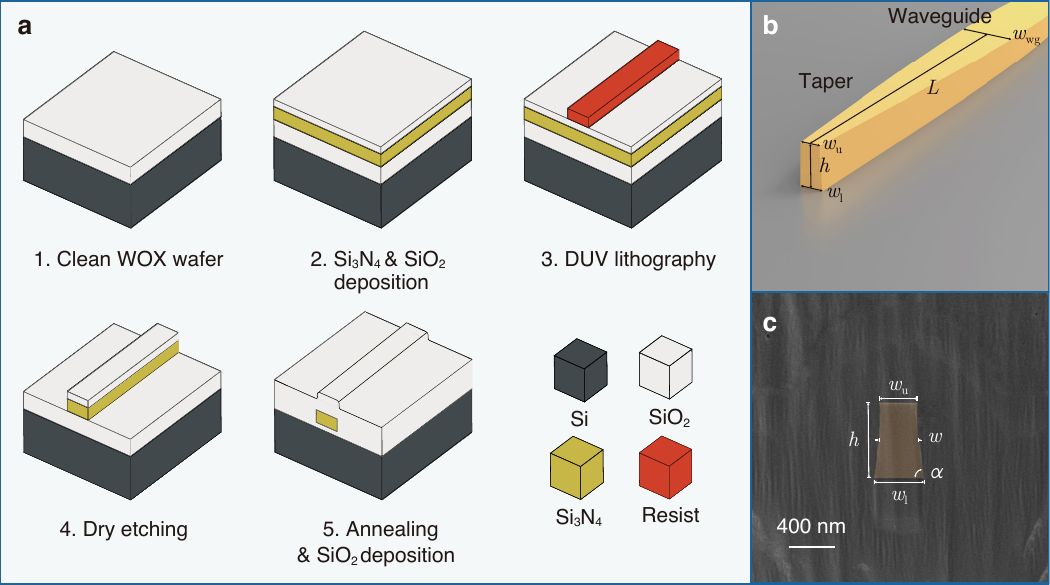}
\caption{
\textbf{Subtractive process flow and illustration of the inverse taper structure.} 
\textbf{a}. 
Simplified Si$_3$N$_4$ subtractive process flow. 
\textbf{b}. 
Illustration of the inverse taper. 
\textbf{c}. 
SEM image of the taper's cross-section at the chip facet. 
}
\label{Fig:5}
\end{figure*}

\begin{figure*}[t!]
\centering
\includegraphics{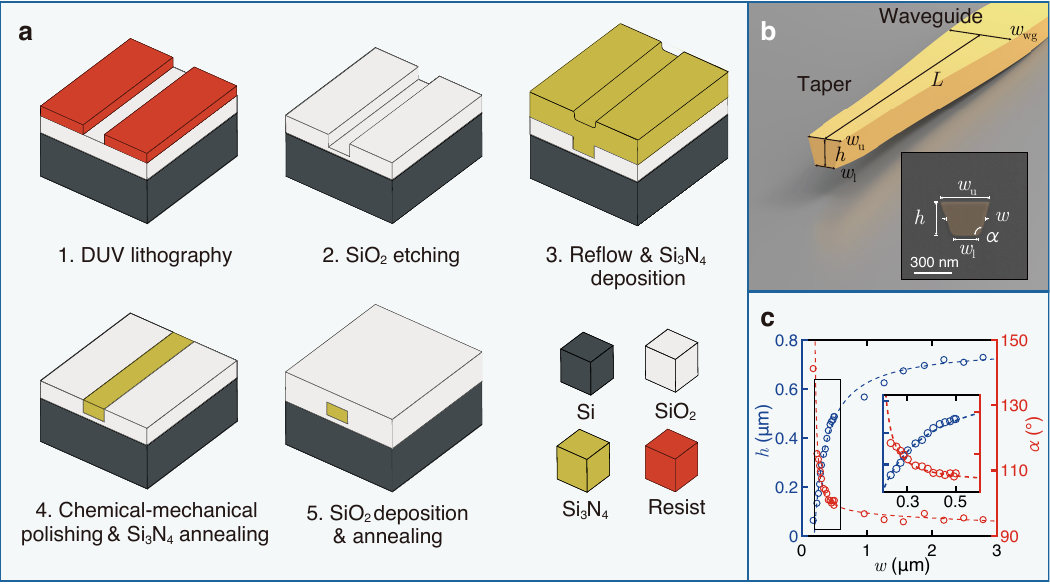}
\caption{
\textbf{Additive process flow and illustration of the 2D-inverse taper structure.} 
\textbf{a}. 
Simplified Si$_3$N$_4$ additive process flow. 
\textbf{b}. 
Illustration of the 2D-inverse taper. 
Inset, SEM image of the taper's cross-section at the chip facet. 
\textbf{c}. 
SEM characterization of the ARDE effect of the 2D-inverse taper. 
The blue circles are measured via SEM, and the blue dashed curve is the ARDE fit.
The red circles are calculated via the SEM characterization result, $w_\text{u}$ and $w_\text{l}$.
The red dashed curve demonstrates the variation of the taper's sidewall angle with width.
}
\label{Fig:6}
\end{figure*}

Mathematically, edge detection relies on calculating the gradient of the gray-scale image.
We accurately segment the gray-scale SEM image into the tip region and the background using the $k$-means clustering algorithm and binarize the two regions\cite{Arthur:06}.
This step enhances the gradient specifically at the tip edge.
Using the function edge in MATLAB, we further extract the curve and remove any anomalous patterns.
Figure \ref{Fig:4}a left shows the gray-scale SEM image of the lensed fiber's tip with $D=3.0\ \mu\text{m}$.
Figure \ref{Fig:4}a right displays a clean and complete gray-scale image of the outline curve, which resembles a hyperbolic shape.
The green and blue dashed lines represent the asymptotes, where $\Theta_l$ and $\Theta_r$ denote the angles between the left and right asymptotes and the $x$ axis, with counter-clockwise angles defined as positive.

Here, we use a hyperbolic function to fit the outline curve of the lensed fiber's tip.
The fiber tip in the SEM image may not be symmetric on the $x$ axis, i.e., $\Theta_l + \Theta_r\neq0$, which complicates the curve fitting.
Therefore, we apply a global rotational correction to the image in Fig. \ref{Fig:4}a right, with the rotation angle given by $\Theta=-(\Theta_l+\Theta_r)/2$.
To determine $\Theta_l$ and $\Theta_r$, we perform a Hough transform on the image in Fig. \ref{Fig:4}a right.
The Hough transform helps identify linear features\cite{Ballard:81}.
For a hyperbolic-shaped curve, two such linear features correspond to its two asymptotes.
As illustrated in Fig. \ref{Fig:4}b, the Hough transform of the image produces two distinct peaks, whose horizontal coordinates correspond to $\Theta_l$ and $\Theta_r$.
We then rotate the coordinate system of the curve by $\Theta$, as
\begin{equation}
    \begin{pmatrix}
    y\\x
    \end{pmatrix}_{\text{rotated}}
=\begin{pmatrix}  
  \cos\Theta & -\sin\Theta \\  
  \sin\Theta & \cos\Theta  
\end{pmatrix} \begin{pmatrix}
    y\\x
    \end{pmatrix}_{\text{raw}}
    \label{Rotation}
\end{equation}
In Fig. \ref{Fig:4}c, the red data represent the original (raw) curve, while the blue data show the curve after rotation.
We perform a second-order polynomial fit on the blue data in the Cartesian coordinate system, as
\begin{equation}
y^2=\sum \limits_{i=0}^{2}q_{i}x^i
\label{polyfit}
\end{equation}
where $q_i$ is the $i$-th order fitting coefficient.
Since the dependent variable in Eq. \ref{polyfit} is $y^2$, the symmetry of the curve data on the $x$ axis avoids errors that would arise from having multiple values $y^2$ corresponding to the same independent variable $x$.
Figure \ref{Fig:4}c shows that the fit hyperbola (green curve) aligns well with the rotated curve (blue data).
After fitting, the parameters $(\rho,\phi)$ of the hyperbola can be derived as
\begin{equation}
\begin{cases}
\rho=-q_1/2 \\
\phi=2\arctan(\sqrt{q_2})
\end{cases}
\label{rho_phi}
\end{equation}

\vspace{0.3cm}
\section*{Appendix B: Subtractive process flow and tapers}
All of the Si$_3$N$_4$ chips used in this work are fabricated in our foundry-level platform. 
Figure \ref{Fig:5}a illustrates the subtractive process flow, starting with a clean, 150-mm-diameter (6-inch), thermal wet oxide (WOX) wafer.
Silicon nitride and SiO$_2$ (as the etch hardmask for Si$_3$N$_4$) are deposited on the WOX wafer via low-pressure chemical vapor deposition (LPCVD). 
The waveguide pattern, defined using a deep-ultraviolet (DUV) scanner lithography (ASML PASS 5500/850C, with 248 nm KrF source and 110 nm lithography resolution), is transferred into the SiO$_2$ hardmask via dry etching. 
After ashing the photoresist, Si$_3$N$_4$ is dry etched with etchants comprising SF$_6$, CHF$_3$, and O$_2$, which creates smooth and vertical sidewalls of the waveguides to minimize optical scattering loss. 
This dry etching step defines all the waveguides geometries, including the taper structure presented in Fig. \ref{Fig:5}b.
Afterwards, the entire wafer is annealed above 1200$^\circ$C to remove residual hydrogen content in Si$_3$N$_4$, followed by LPCVD SiO$_2$ cladding deposition and SiO$_2$ annealing again above 1200$^\circ$C. 
Finally, chip facets are defined and created by UV lithography and deep SiO$_2$ and Si dry etching. 
The entire wafer is separated into hundreds of chips by dicing or backside grinding. 

Figure \ref{Fig:5}c shows the SEM image of the taper's cross-section at the chip facet.
The parameters, $w_\text{u}$, $w_\text{l}$, and $h$, are measured by the SEM (CIQTEK SEM5000Pro).
We approximate the taper's cross-section to a simple trapezoid.
Then $w$ and $\alpha$ are defined as
\begin{equation}
    \begin{cases}
        w=\frac{w_\text{u}+w_\text{l}}{2}\\
        \alpha=\frac{\pi}2+\arctan{(\frac{w_\text{u}-w_\text{l}}{2h}})
    \end{cases}
    \label{w_alpha}
\end{equation}
where $\alpha$ is independent of $w$ for the subtractive process.

\vspace{0.3cm}
\section*{Appendix C: Additive process flow and tapers}
Figure \ref{Fig:6}a illustrates the additive process flow, starting with a clean, 6-inch, WOX wafer. 
Waveguide preform, as well as the filler pattern to release the Si$_3$N$_4$ film's high tensile stress \cite{Pfeiffer:18b}, are defined by DUV scanner lithography. 
These patterns are subsequently transferred from the photoresist mask into the WOX via dry etching. 
After ashing the photoresist, a preform reflow at 1250$^\circ$C is performed to reduce the preform's sidewall roughness \cite{Pfeiffer:18}, thereby minimizing optical Rayleigh scattering loss. 
A thick Si$_3$N$_4$ film is deposited onto the patterned wafer via LPCVD, filling the trenches and forming the waveguides.
Chemical mechanical polishing (CMP) follows to remove excess Si$_3$N$_4$, leaving a planarized wafer surface with Si$_3$N$_4$ waveguides embedded  in SiO$_2$. 
Afterwards, the entire wafer is annealed above 1200$^\circ$C to remove residual hydrogen content in Si$_3$N$_4$, followed by LPCVD SiO$_2$ cladding deposition and SiO$_2$ annealing again above 1200$^\circ$. 
Finally, chip facets are defined and created by UV lithography and deep SiO$_2$ and Si dry etching. 
The entire wafer is separated into hundreds of chips by dicing or backside grinding. 

This process creates the 2D-inverse tapers shown in Fig. \ref{Fig:6}b.
The taper geometry is primarily determined by the SiO$_2$ preform dry etching and chemical mechanical polishing (CMP) methods described in Ref. \cite{Liu:18}.
We characterize the taper's geometry at chip facets via SEM. 
The parameters, $w_\text{u}$, $w_\text{l}$, and $h$, are measured via SEM, while $w$ and $\alpha$ are calculated with Eq. \ref{w_alpha}.  
Figure \ref{Fig:6}c illustrates the aspect-ratio dependent etch (ARDE) effect, causing tapers with varying $w$ and $h$, different from the subtractive tapers of varying $w$ and constant $h$. 
The ARDE effect, detailed in Ref. \cite{Gottscho:92, Liu:18}, refers to the dimension-dependent etch rate in high-aspect-ratio structures, particularly prominent for narrow trenches.   
A narrower trench exhibits a lower etch rate because etch occurs less efficiently in deep, narrow spaces, while reaction by-products are more difficult to remove. 
This leads to a reduced etch rate and altered etch profiles in narrow tapers. 

We systematically optimize the etch pressure to control the degree of ARDE, thereby fabricating 2D-inverse tapers. 
The non-uniform etch resulting from the ARDE effect and its dependence on pattern width are directly observable in SEM. 
This dependence is shown in Fig. \ref{Fig:6}c, indicating a reduction in trench height for a width below 500 nm.
It is fitted with Eq.~\ref{FIT}, and the fit results are illustrated with blue and red dashed curves.
The fit results are close to the data, indicating that the trend of ($h$, $\alpha$) approaches (770 nm, 90$^\circ$) as $w$ increases.

\vspace{0.3cm}
\section*{Appendix D: FDTD simulation}
\begin{figure}[htbp]
\centering
\includegraphics[width=0.5\textwidth]{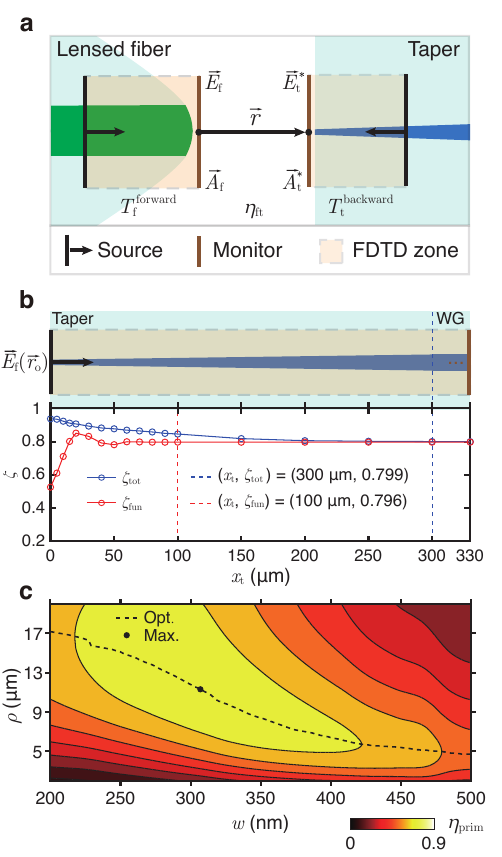}
\caption{
\textbf{FDTD simulation.} 
\textbf{a.}  
Schematic illustration of the coupling process and the computational framework for the overlap integral.
The fields ($\vec{E}_{\text{f}}$ and $\vec{E}_{\text{t}}$), and their associated transmission coefficients ($T^{\text{forward}}_{\text{f}}$ and $T^{\text{backward}}_{\text{t}}$) derived from 3D-FDTD simulations, serve as inputs for the calculation defined in Eq. \ref{eta_ft}.
\textbf{b.} 
Calculated transmission coefficients for a $D=6~\mu\text{m}$ lensed fiber coupling to a subtractive taper with geometry ($w, h, \alpha$) = (170 nm, 320 nm, 84$^\circ$) under TE polarization. 
$\zeta(x_\text{t})$ denotes the power transmission from the incident field $\vec{E}_{\text{t}}(\vec{r}_o)$ at the facet to an arbitrary cross-section at a relative distance $x_\text{t}$. 
$\zeta_{\text{tot}}$ represents the total power transmission, while $\zeta_{\text{fun}}$ corresponds to the fundamental mode transmission.
\textbf{c.} 
Coupling efficiency $\eta_{\text{prim}}$ (indicated by the colorbar) for a $\phi=90^\circ$ lensed fiber and a ($h, \alpha$) = (320 nm, 84$^\circ$) subtractive taper under TE polarization at 1550 nm wavelength, computed via Eq. \ref{eta_prim} across a parameter space of ($\rho, w$).
It is a six-layer contour plot, presented to highlight both the uniqueness and visual appeal of the optimal solution.
The dashed line (Opt.) traces the optimal $\rho$ for any given $w$.
The marker (Max.) indicates the global optimum at ($\rho, w, \eta_{\text{prim}})_{\text{Max.}}=(11~\mu\text{m}, 310~\text{nm}, 0.777)$.
}
\label{Fig:7}
\end{figure}

The first step to simulate light coupling efficiency from the lensed fiber to the inverse taper is to calculate the optimal relative position $\vec{r}_o$ of the lensed fiber and taper. 
The value of $\vec{r}_o$ is defined as the position that maximizes the overlap integral given by Eq. \ref{OI}. 
The field profiles $\vec{E}_\text{f}$ and $\vec{E}_\text{t}$ required for this calculation are acquired via FDTD simulation.
The simulation configuration is shown in Fig. \ref{Fig:7}a. 
For the lensed fiber, the fundamental mode is excited and a monitor at the tip collects the output field $\vec{E}_\text{f}$ and measures the transmittance $T^{\text{forward}}_{\text{f}}$. 
The angular spectrum of this field $\vec{A}_\text{f}$ can be calculated by the Fourier transform:
\begin{equation}
\begin{split}
\vec{A}_{\text{f}}(x,k_y,k_z)=\mathscr{F}[\vec{E}_{\text{f}}(x,y,z)] \\
=\frac{1}{4\pi^2}\iint \vec{E}_{\text{f}}(x,y,z)e^{- i(k_y y+k_z z)}\mathrm{d}y\mathrm{d}z
\end{split}
\label{fourier}
\end{equation}
For the taper, an equivalently polarized fundamental mode is excited, with a monitor at the other end that captures its field $\vec{E}^\ast_{\text{t}}$ and transmittance $T^{\text{backward}}_{\text{t}}$.
Similarly, we can calculate its angular spectrum $\vec{A}^\ast_\text{t}$ using Eq. \ref{fourier}.
The propagation of light in air is governed by the Helmholtz propagator\cite{Matsushima:09}, which allows the output field after a propagation vector $\vec{r}=(x,y,z)$ to be derived from the angular spectrum as
\begin{equation}
\vec{A}_\text{f}(\vec{r})=\vec{A}_\text{f}\cdot e^{i\vec{k}\cdot\vec{r}}
\label{angle_spectrum_method}
\end{equation}
where $\vec{k}$ is determined by
$\vec{k}=(k_x,k_y,k_z)$ and satisfies $|\vec{k}|=\frac{2\pi}{\lambda}$, $\vec{A}_\text{f}(\vec{r})$ is the angular spectrum at $\vec{r}$.
The corresponding field $\vec{E}_{\text{f}}(\vec{r})$ can be calculated using an inverse Fourier transform according to Eq. \ref{fourier}.
Furthermore, based on Parseval's theorem and Eqs. \ref{fourier} and \ref{angle_spectrum_method}, the integral of the electric field intensity over space can be expressed as
\begin{align}
\iint \vec{E}_\text{f}(\vec{r})\cdot \vec{E}^\ast_\text{t} \ \mathrm{d}y\mathrm{d}z=\frac{1}{4\pi^2}\iint \vec{A}_\text{f}(\vec{r})\cdot \vec{A}_\text{t}^\ast\ \mathrm{d}k_y\mathrm{d}k_z \label{parseval_1}\\
\iint |\vec{E}_\text{f}(\vec{r})|^2 \ \mathrm{d}y\mathrm{d}z=\frac{1}{4\pi^2}\iint |\vec{A}_\text{f}|^2\ \mathrm{d}k_y\mathrm{d}k_z \label{parseval_2}\\
\iint |\vec{E}_\text{t}|^2 \ \mathrm{d}y\mathrm{d}z=\frac{1}{4\pi^2}\iint |\vec{A}_\text{t}|^2\ \mathrm{d}k_y\mathrm{d}k_z \label{parseval_3}
\end{align}
where Eq. \ref{parseval_2} embodies the conservation of optical energy in propagation through the air.
Substituting Eqs. \ref{angle_spectrum_method} and \ref{parseval_3} into Eq. \ref{OI} yields
\begin{equation}
\eta_\text{ft}(\vec{r})=\frac{\left|\iint  \vec{A_{\text{f}}} \cdot \vec{A_{\text{t}}}^{\ast}\cdot e^{i\vec{k}\cdot \vec{r}}\  \mathrm{d}k_y\mathrm{d}k_z\right|^{2}}{\iint\left|\vec{A_{\text{f}}}\right|^{2} \mathrm{d}k_y\mathrm{d}k_z  \cdot \iint\left|\vec{A_{\text{t}}}\right|^{2} \mathrm{d}k_y\mathrm{d}k_z}
\label{eta_ft}
\end{equation}

The final determination of $\vec{r}_o$ is achieved by this formula coupled with a simulated annealing algorithm.
This method is advantageous for two main reasons.
First, it avoids computationally expensive iterations of
$\vec{E}_\text{f}(\vec{r})$, leading to a substantial reduction in optimization time. 
Second, the methodology can be extended to optimize the coupling between arbitrary pairs of couplers. 
Provided that the taper is reciprocal ($T^{\text{forward}}_\text{t}=T^{\text{backward}}_\text{t}$), a preliminary estimate of the coupling efficiency $\eta_{\text{prim}}$ given by 
\begin{equation}
\eta_{\text{prim}}=T^{\text{forward}}_\text{f}\cdot\eta_{\text{ft}}(\vec{r}_o)\cdot T^{\text{backward}}_\text{t}
\label{eta_prim}
\end{equation}
This approach enables rapid optimization of key design parameters for both couplers.
As demonstrated in Fig. \ref{Fig:7}c for a lensed fiber with $\phi=90$$^\circ$ and a subtractive taper with $h=320$ nm and $\alpha=84$$^\circ$, the optimized parameters $(\rho,w)$ = (11 $\mu$m, 310 nm) achieve a maximum $\eta_{\text{prim}}=77.7\%$, which is consistent with our experimental results.
The dashed line, indicating the optimal pair ($\rho$, $w$) for certain $w$, exemplifies the optimization strategy for the lensed fiber, which is experimentally validated in this work.

However, the general applicability of coupler reciprocity may not be assured, and complex near-field interactions are elusive.
Therefore, we proceed with a 3D FDTD simulation to determine the transmission efficiency $\zeta$ from the taper facet to the waveguide's mode.
The beam profile at $\vec{r}_o$ (i.e., $\vec{E}_{\text{f}}(\vec{r}_o)$) is imported as a source incident on the end face of the taper waveguide.
As shown in Fig. \ref{Fig:7}b, since the fundamental mode in the waveguide taper becomes stable after propagating approximately 100 $\mu$m--for a total taper length of 300 $\mu$m--we place a mode expansion monitor 100 $\mu$m from the taper's input facet.
The transmission efficiency of the fundamental mode, indicated as $\zeta_{\text{fun}}$, is collected on this monitor.
The total coupling efficiency is then calculated as the product of these two values.
\begin{equation}
\eta_{\text{sim}}=T^{\text{forward}}_\text{t}\cdot \zeta_{\text{fun}}
\label{eta_sim}
\end{equation}
This allows us to truncate the simulation domain, reducing its size by more than two-thirds.

\vspace{0.3cm}
\begin{figure*}[t!]
\centering
\includegraphics[width=\textwidth]{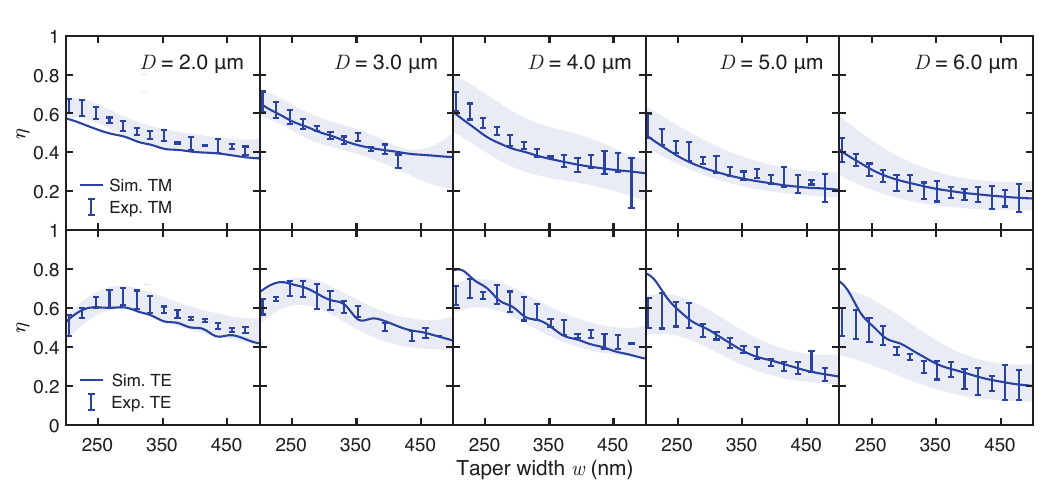}
\caption{
\textbf{Supplementary experimental characterization of light coupling efficiency $\eta$.} 
Experimentally (Exp.) characterized $\eta$ for 830-nm-thick subtractive tapers with varying $D$ and $w$ under TE or TM polarization, in comparison with simulation (Sim.) results.
}
\label{Fig:9}
\end{figure*}
\begin{figure}[h]
\centering
\includegraphics[width=0.5\textwidth]{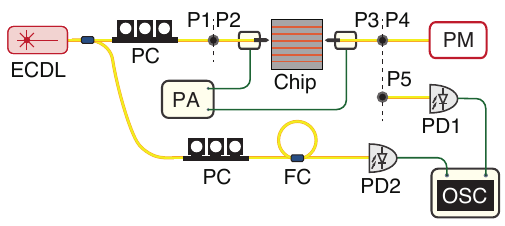}
\caption{
\textbf{Experimental setup.} 
Schematic of the experimental setup for characterizing coupling efficiency, misalignment tolerance, and transmission spectrum.
The chip contains many inverse tapers and waveguides with various taper widths.
ECDL, external-cavity diode laser.
PC, polarization controller.
PM, optical power meter.
PA, piezoelectric actuator.
FC, fiber cavity.
PD1 and PD2, photodetectors at the test and reference ports, respectively.
OSC, oscilloscope.
}
\label{Fig:8}
\end{figure}
\section*{Appendix E: Experimental setup}

The single-facet efficiency $\eta$ is determined as the square root of the total coupling efficiency measured through both ends, i.e. fiber-chip-fiber.
As depicted in Fig. \ref{Fig:8}, we measure the $\eta$ by first connecting P1 to P2 and P3 to P4 via flanges.
After optimizing the coupling position, the laser wavelength is tuned within a range of approximately $\pm$ 0.12 nm (corresponding to $\pm$15 GHz) around 1550 nm, and the maximum efficiency value within this range is taken as the final result for each facet to minimize the influence of facet reflections (i.e. the visible FP interference in Fig.~\ref{Fig:3}c).
Then the output power $P^{\text{optimal}}_{\text{out}}$ is measured with a tunable laser (Santec TSL570) and a power meter (Thorlabs 100D).
We use a piezo driver (Thorlabs MDT694B) to scan the actuators (Thorlabs Nanomax300) in the 20 $\mu$m range to perform a coupling misalignment test, as shown in Fig.~\ref{Fig:3}b.
The reference input power $P_{\text{in}}$ is measured by connecting P1 to P4 under the same laser conditions, allowing the calculation of $\eta$ as 
\begin{equation}
\eta=\sqrt\frac{P^{\text{optimal}}_{\text{out}}}{P_{\text{in}}}
\label{eta}
\end{equation}
We measured three taper samples with identical design parameters for each one. 
The data points in Fig.~\ref{Fig:3}a, d represent the average $\eta$ values, with the error bars indicating twice the standard deviation from the three measurements. 
The vertical range of the shaded area indicates the 95\% confidence interval for the coupling efficiency, as predicted by the Gaussian process regression model\cite{Wangj:23}.
For transmission spectrum characterization as shown in Fig.~\ref{Fig:3}c, the setup is reconfigured as in Ref.~\cite{Luo:24} by connecting P1 to P2 and P3 to P5, giving $P_{\text{out}}(\lambda)$.
The reference $P_{\text{in}}(\lambda)$ is measured by connecting P1 to P4, where the transmission spectrum $T(\lambda)=\frac{P_{\text{out}}(\lambda)}{P_{\text{in}}(\lambda)}$ is derived.
\vspace{0.3cm}
\section*{Appendix F: Experimental Results for 830-nm-thickness subtractive tapers}
Figure~\ref{Fig:9} shows the measured $\eta$ with varying $w$, for 830-nm-thick subtractive tapers.
The experimental data also agree well with the FDTD simulation results. 
As can be seen in Figure~\ref{Fig:3}a, for tapers fabricated using the same subtractive process, different taper thickness correspond to different optimal taper widths $w$ under the same $D$ and polarization mode for lensed fiber coupling.
The optimal $w$ for 830 nm tapers is generally smaller than that for 320 nm tapers.
When the corresponding $D$ is too large, the optimal taper $w$ exhibits values below the minimum
dimension (160 nm) of our photolithography.

\vspace{0.3cm}
\bibliographystyle{apsrev4-1}
\bibliography{bibliography_cdk_v1}

\end{document}